\documentclass{article}

\usepackage{arxiv}

\usepackage[utf8]{inputenc} 
\usepackage[T1]{fontenc}    
\usepackage{hyperref}       
\usepackage{url}            
\usepackage{booktabs}       
\usepackage{amsfonts}       
\usepackage{nicefrac}       
\usepackage{microtype}      
\usepackage{lipsum}
\usepackage{color}
\usepackage{algorithm}
\usepackage{algorithmic}
\usepackage{amsmath,amsfonts,amssymb}
\usepackage{subfigure}
\usepackage{lineno}
\usepackage{graphicx}
\usepackage{bm}
\usepackage{subfigure}

\title{Data-driven modeling and decomposition for nanoscale liquid-film dynamics: Application to superspreading nanofluid droplets}

\author{
  Kai Fukami\\
  Department of Aerospace Engineering\\
  Graduate School of Engineering\\
  Tohoku University\\
  Sendai, 980-8579, Japan\\
  \texttt{kfukami1@tohoku.ac.jp}
  \And
  Eita Shoji\\
  Department of Mechanical Systems Engineering\\
  Graduate School of Engineering\\
  Tohoku University\\
  Sendai, 980-8579, Japan\\
  \texttt{eita.shoji@tohoku.ac.jp}
}

\begin{document}
\maketitle

\begin{abstract}

Understanding ultrathin liquid-film dynamics is crucial for unraveling complex interfacial phenomena, yet deriving governing equations directly from experimental observations remains challenging. This study proposes a data-driven approach to model droplet dynamics, capturing liquid-film thickness on the nanometer scale in the form of a partial differential equation. 
As a challenging test case, we examine the superspreading wetting of surfactant-free nanofluids, a phenomenon whose physical mechanism defies standard theoretical explanations. We apply a sparse identification algorithm to spatiotemporal film-thickness profiles resolved at the nanometer scale using phase-shifting imaging ellipsometry. For a pure solvent, the discovered governing equation recovers classical lubrication physics driven by disjoining pressure and evaporation. 
In contrast, the nanofluid dynamics necessitates an additional, unique transport term scaling with the gradient of the inverse film thickness. Theoretical scaling analysis suggests this term represents a nanoparticle-induced bias flux, consistent with a hypothesized capillary wicking mechanism within the precursor film. 
The identification of the current nanofluid-specific term underscores the efficacy of integrating high-precision experimental measurements with data-driven modeling to discover hidden physics and generate testable hypotheses in complex wetting dynamics.

\end{abstract}

\section{Introduction}
\label{sec:intro}

The wetting dynamics of complex fluids, such as surfactant solutions and nanoparticle suspensions, play a critical role in various scientific and industrial applications \cite{degennes1985, oron1997-rmp, sefiane2008-acis, bonn2009-rmp, lu2016-acis, lohse2022-arfm}. In many anomalous wetting phenomena, the macroscopic spreading behavior is closely interlinked with the physics within the liquid film near the contact line. For example, specific microstructures and particle arrangements formed within the thin film—such as the stratification of nanoparticles leading to structural disjoining pressure \cite{wasan2003-nature}, bilayer formations in surfactant-laden drops \cite{karapetsas2011-jfm, theodorakis2015-langmuir}, and nanoparticle structuring within the nanoliquid film governing wetting dynamics and deposition patterns \cite{shoji2025-jcis}—are known to significantly modulate the macroscopic spreading behavior. Particularly, the nanometric precursor film advancing ahead of the macroscopic contact line often dictates the fundamental spreading kinetics. While classical theories provide a robust framework for understanding the precursor films of pure, simple liquids \cite{degennes1985,joanny1986}, constructing a model that captures the local dynamics of complex fluids remains a significant challenge.

A prominent example of such challenging phenomena is superspreading wetting, originally discovered by Schwarz and Reid \cite{schwarz1964surface}. Characterized by significant deviations from the simple scaling law between contact radius and time \cite{tanner1979spreading}, the universal mechanisms driving superspreading remain elusive and highly material-dependent \cite{venzmer2021-acis}. Beyond traditional surfactant-based systems, our recent experiments demonstrated that surfactant-free nanofluids composed of highly dispersed single-nanometer particles also exhibit striking superspreading \cite{shoji2024-langmuir}. Crucially, our comprehensive characterizations have systematically ruled out conventional mechanisms.
{As detailed in our previous work \cite{shoji2024-langmuir}, surface-tension measurements did not support classical surfactant-driven Marangoni stresses as the dominant driving mechanism \cite{nikolov2002-acis,williams2020-jfm}; time-resolved thin-film profiling did not show the characteristic behavior expected for a structural-disjoining-pressure mechanism based on nanoparticle stratification or layering \cite{wasan2003-nature}; and the observed shear-thickening rheology \cite{hossain2019} is inconsistent with spreading models relying on viscosity reduction.}
{These observations indicate that conventional mechanisms are insufficient to explain the present superspreading dynamics.}

{At the same time, these observations do not imply that nanoparticles are irrelevant; rather, they motivate a closer examination of nanoparticle-induced transport within the nanometric precursor film.}
{Our previous post-drying observations suggested that nanoparticles can persist within the precursor-film region \cite{shoji2024-langmuir}.}
{Subsequent experiments under evaporative inkjet-like conditions, in which droplet evolution from deposition to drying was continuously monitored by phase-shifting imaging ellipsometry, further indicated the presence of nanoparticles in the nanoliquid-film region and showed that nanofluid droplets can spread to a larger maximum radius than the corresponding base liquid \cite{shoji2025-jcis}.}
{More recently, direct measurements of nanoparticle concentration distributions suggested that superspreading nanofluids exhibit nanoparticle enrichment near the contact line \cite{shoji2026-ami}.}
{Despite these experimental indications of the importance of nanoparticles within the nanometric film, formulating a local governing equation for such complex nanofluid wetting dynamics remains difficult because the relevant transport mechanisms are not known a priori.}

{To convert these experimental indications into a mechanistic description, it is necessary not only to observe the nanometric film but also to extract local dynamical laws from its spatiotemporal evolution.} While techniques such as atomic force microscopy \cite{villette1997-pa,xu2004-prl,glynos2011-prl} and X-ray reflectivity \cite{heslot1989-jpcm, cazabat1991-acis} have provided valuable insights into static or slow-moving thin films, capturing the transient evolution of nanoscale precursor films remains experimentally challenging. {Phase-shifting imaging ellipsometry overcomes this limitation by resolving nanometer-thick liquid films during dynamic wetting, as demonstrated in our previous studies \cite{shoji2025-jcis,shoji2024-langmuir,shoji2026-ami,shoji2019-olen,shoji2021-ef}. However, deducing a governing equation or identifying the operative transport mechanism solely from such measured kinematic profiles remains difficult. To address this remaining challenge, we combine these nanometer-resolved film-thickness data with a data-driven sparse modeling approach, PDE-FIND \cite{rudy2017data}.} Equipped with sparsity-promoting optimizations, the algorithm discovers a modeled partial differential equation from time-discretized data by identifying dominant terms from a predefined library of potential physical operators \cite{brunton2016discovering}. Combined with other data-driven techniques, such as nonlinear machine-learning-based reduced-order modeling \cite{champion2019data,fukami2020sparse}, such sparsity-promoting approaches have been successfully used for the prediction and control of a range of scientific and engineering problems, including robotics \cite{chu2020discovering}, biology \cite{mangan2017inferring}, solid mechanics \cite{klishin2025statistical}, and unsteady aerodynamics \cite{scherl2020robust,hickner2023data}. {To the best of our knowledge, however, this is the first application of such a sparse equation-discovery technique to experimentally measured nanoscale liquid-film dynamics, particularly to nanometric precursor-film profiles in complex wetting.}

{The specific research gap addressed in this study is therefore the lack of a local, experimentally grounded governing-equation description for surfactant-free nanofluid superspreading after conventional mechanisms have been found insufficient.
The objective of this work is not to assume a particular microscopic mechanism a priori, but to identify sparse PDE operators governing the measured film-thickness dynamics directly from nanometer-resolved profiles.
In this strategy, the pure liquid ($n$-heptane) case serves as a validation case in which the data-driven method is expected to recover classical lubrication-type physics, whereas the CeO$_2$ nanofluid case is used to determine whether an additional nanofluid-specific transport operator is required.
By combining high-resolution phase-shifting imaging ellipsometry with sparse PDE discovery, we aim to extract both a compact mathematical representation of the local thin-film dynamics and a physically interpretable hypothesis for the anomalous superspreading mechanism.
Such mechanistic insight may also provide a basis for future continuum modeling of wetting processes involving highly dispersed nanoparticles, although the present study focuses on the fundamental nanoscale transport mechanism rather than on direct process optimization.}

This paper is organized as follows. The approach is described in section \ref{sec:method}. Results of model discovery for the nanometer-thick liquid film data sets are presented in section \ref{sec:res}. Conclusions are offered in section \ref{sec:conc}.

\begin{figure}
    \centering
    \includegraphics[width=1\textwidth]{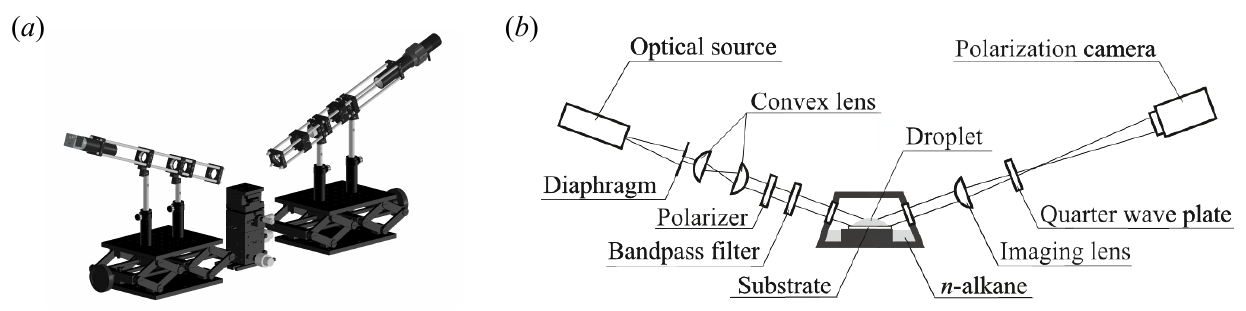}
    \vspace{-5mm}
    \caption{
    The phase-shifting imaging ellipsometer and sealed environmental cell used in the present analysis: $(a)$ the model and $(b)$ detailed descriptions are shown. 
    The optical configuration and calibration/uncertainty procedures are identical to our prior studies~\cite{shoji2019-olen,shoji2024-langmuir}.}
    \vspace{-3mm}
    \label{fig1}
\end{figure}

\section{Approach}
\label{sec:method}

{
This study aims to extract physical insights from experimental data on superspreading nanofluids through sparse dynamical modeling.
We consider previously published experimental data~\cite{shoji2024-langmuir}, acquired via phase-shifting imaging ellipsometry shown in figure~\ref{fig1}.
The capability of this technique to resolve nanoscale precursor film dynamics has been demonstrated across various wetting phenomena, ranging from particle suspensions~\cite{shoji2021-ef} to evaporating nanofluids~\cite{shoji2025-jcis}.
This setup enables measuring spatiotemporal thickness fields from nanometric to micrometric scales.
We examine two liquids: pure $n$-heptane and a surfactant-free 5~wt\% nanofluid containing surface-modified CeO$_2$ nanoparticles dispersed in $n$-heptane.
Silicon wafers served as substrates, with surface preparation and optical modeling.
The environmental cell was saturated with the working liquid to suppress evaporation.
}

\begin{figure}[t]
    \centering
    \includegraphics[width=0.87\textwidth]{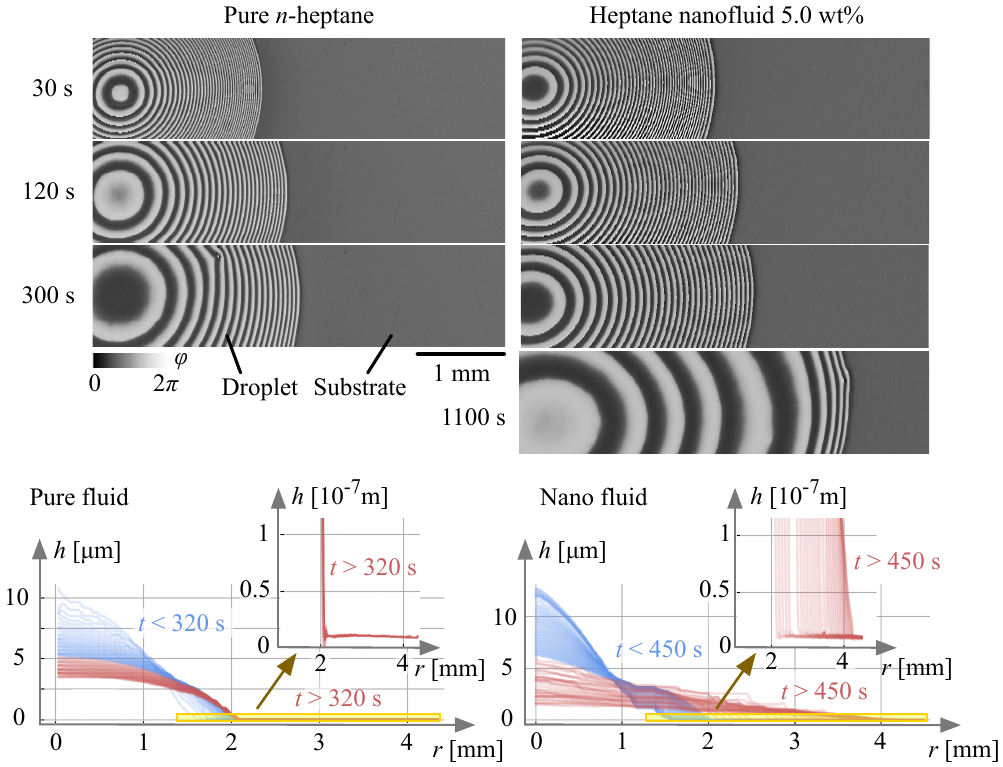}
    \caption{
    Phase difference field $\varphi$ of the pure $n$-heptane and the heptane 5~wt\% nanofluid along with their height dynamics.
    The zoomed-in view of the region of interest for the current data-driven analysis, highlighted with yellow boxes, is also shown.  
    }
    \label{fig2}
\end{figure}

{
Representative flow snapshots visualized as the phase difference distribution $\varphi$ between $p$- and $s$-polarizations are presented in figure~\ref{fig2}, where the observed interference fringes serve as topographic contours of the liquid film.
Along with these snapshots, the corresponding radial thickness profiles $h(r,t)$ are plotted.
The insets in the thickness plots highlight the dynamics within the nanometric precursor region ($h \lesssim 10^{-7}\,\mathrm{m}$), specifically focusing on the timeframes where the distinct behaviors become most pronounced: $t > 320$~s for the pure liquid, where spreading is arrested by evaporation, and $t > 450$~s for the nanofluid, where superspreading becomes dominant.
This pairing serves as a testbed for data-driven discovery. 
That is, the pure liquid validates the method against classical lubrication physics, while the nanofluid presents an unexplained dynamical anomaly.
Crucially, our analysis targets this nanometric precursor region ($h \lesssim 100\,\mathrm{nm}$).
Since our previous post-drying observations suggested that nanoparticles persist even within this precursor film region, we posit that the singular physics in this $O(10^{-7})\,\mathrm{m}$ vicinity, governed by disjoining pressure and particle-induced effects, ultimately dictate the macroscopic spreading dynamics.
}

\begin{figure}
    \centering
    \includegraphics[width=1.05\textwidth]{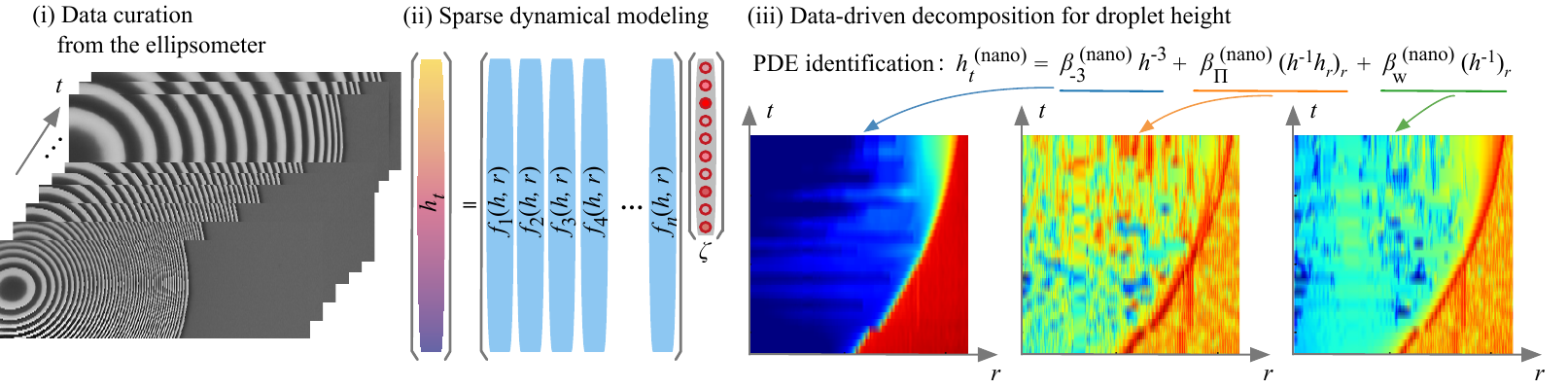}
    \vspace{-5mm}
    \caption{
    Overview of the present analysis: (i) Data curation from the ellipsometer, (ii) sparse dynamical modeling, and (iii) data-driven decomposition for the droplet height based on the modeled equations.}
    \vspace{-3mm}
    \label{fig3}
\end{figure}

{
Equipped with the superspreading droplet datasets, we aim to identify a model equation of height dynamics as a form of partial differential equation, i.e., $h_t = \mathcal{N}(h, h_r, h_{rr}, \dots)$, where the temporal evolution of the height profile is expressed as a linear combination of nonlinear spatial derivatives, as summarized in figure~\ref{fig3}.
This is achieved with a data-driven technique, PDE-FIND~\cite{rudy2017data}, enabling the extraction of model equations from given time-series data in a sparse, parsimonious form.
The PDE-FIND models the time-varying height dynamics as the superposition of candidate physical terms, referred to as the library matrix ${\bm \Theta}(h)$ detailed later, with the weighting coefficients $\bm \beta$, such that ${\bm H}_t = {\bm \Theta}(h){\bm \beta}$.
Given the discretized height data $h(r,t)$ measured at $m$ spatial points and $n$ temporal points, ${\bm H}_t \in \mathbb{R}^{mn}$ is the column vector of the temporal derivatives of height, and ${\bm\beta} \in \mathbb{R}^D$ is the vector of coefficients for $D$ library terms.
Once the droplet dynamics is identified as a PDE form, it is also possible to examine the spatiotemporal pattern of each term appearing in the modeled equation, enabling us to capture the transient characteristics in a decomposed manner.

While generic polynomial expansions are often considered for library construction~\cite{brunton2016discovering,rudy2017data}, this study considers populating them based on our prior knowledge of droplet dynamics, derived from the lubrication approximation, segregated into regions dominated by intermolecular forces and those dominated by surface tension.
We hereafter use $h_t\equiv \partial h/\partial t$ and $h_r\equiv \partial h/\partial r$ for compactness.
}

\begin{table}[htbp]
\centering
\caption{
{Library candidate terms used for the present data-driven analysis.
}}
\label{tab:SINDycandidate}
\begin{tabular}{ll}
\toprule
\textbf{Candidate Term} & \textbf{Classification / Physical Interpretation} \\ 
\midrule
$1$                     & Baseline constant term \\
$h^{-3}$                & Canonical disjoining pressure scale ($\Pi$) \\
\midrule
$(h^{-1}h_r)_r$         & Divergence of conservative flux (non-retarded van der Waals) \\
$(h^{-1})_r$            & Thinness-activated transport candidate \\
$(h^{-2})_r$            & Thinness-activated transport candidate \\
$(h^3\Pi_r)_r$          & Alternative driving force formulation ($\Pi \propto h^{-3}$) \\
\midrule
$h^{-1} h_{rr}$         & Expanded non-conservative derivative \\
$h^{-2} (h_r)^2$        & Expanded non-conservative derivative \\
\midrule
$(h^{-2}h_r)_r$         & Conservative neighbor: weak-slip-like mobility \\
$(h^{0}h_r)_r$          & Conservative neighbor: catch-all \\
\bottomrule
\end{tabular}
\end{table}

Considering the region of interest, the precursor region where disjoining pressure drives the flow, the library prioritizes terms scaling with negative powers of the film height.
Specifically, we incorporate the divergence of the conservative flux for non-retarded van der Waals forces, $(h^{-1}h_r)_r$, alongside candidate terms for thinness-activated transport, such as $(h^{-1})_r$ and $(h^{-2})_r$.
To enhance robustness against measurement noise and discretization errors, we also include expanded non-conservative derivatives, e.g., $h^{-1} h_{rr}$, and alternative formulations of the driving force, such as $(h^3 \Pi_r)_r$, where $\Pi \propto h^{-3}$ represents the canonical disjoining pressure.
The resulting library matrix ${\bm \Theta}(h)$ is expressed as
\begin{equation}
{\bm \Theta}(h) = \left[ \ 
1, \ \ {h^{-3}, \ \ (h^{-1} h_r)_r, \ \ (h^{-1})_r, \ \ (h^{-2})_r, \ \dots} \ \
{h^{-1} h_{rr}, \ \ (h^3\Pi_r)_r} 
\ \right],
\end{equation}
{where the full library candidates are showcased in table~\ref{tab:SINDycandidate}.}
Once the library matrix ${\bm \Theta}(h)$ is defined, the coefficients for each term in the library are sought through the iterative optimization,
\begin{align}
    {\bm \beta}^* = {\rm argmin}_{\bm \beta}||{\bm \Theta} {\bm \beta}-{\bm H}_t||_2 + \alpha ||{\bm \beta}||_1,
    \label{eq:SINDy}
\end{align}
where $(\cdot)^*$ represents the optimized variable.
Here, the coefficients are determined through the sequential threshold least squares method~\cite{brunton2016discovering}, promoting sparsity of the coefficient matrix in a computationally tractable manner.
In performing the optimization of equation \ref{eq:SINDy}, all the variables are standardized with their mean and standard deviation, and then rescaled back to the original dimension.

\section{Results}
\label{sec:res}
{\subsection{Sparse identification and reconstruction of film-thickness dynamics}}

\begin{figure}
    \centering
    \includegraphics[width=0.65\textwidth]{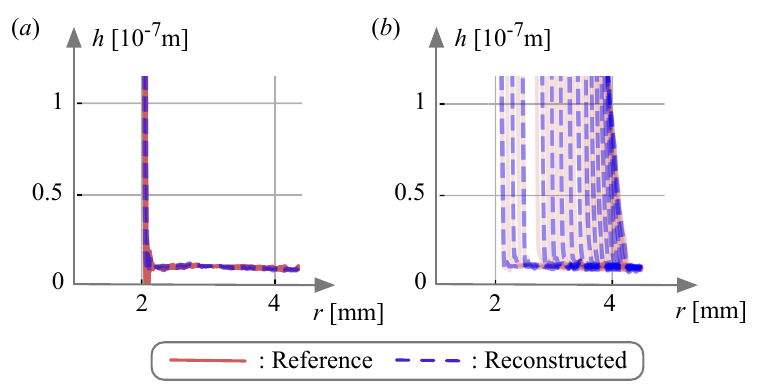}
    \caption{{
    Comparison between the evolutions of the measured and reconstructed film thickness for $(a)$ pure and $(b)$ nano fluids.
    }}
    \label{fig:rec}
\end{figure}

\begin{figure}
    \centering
    \includegraphics[width=0.9\textwidth]{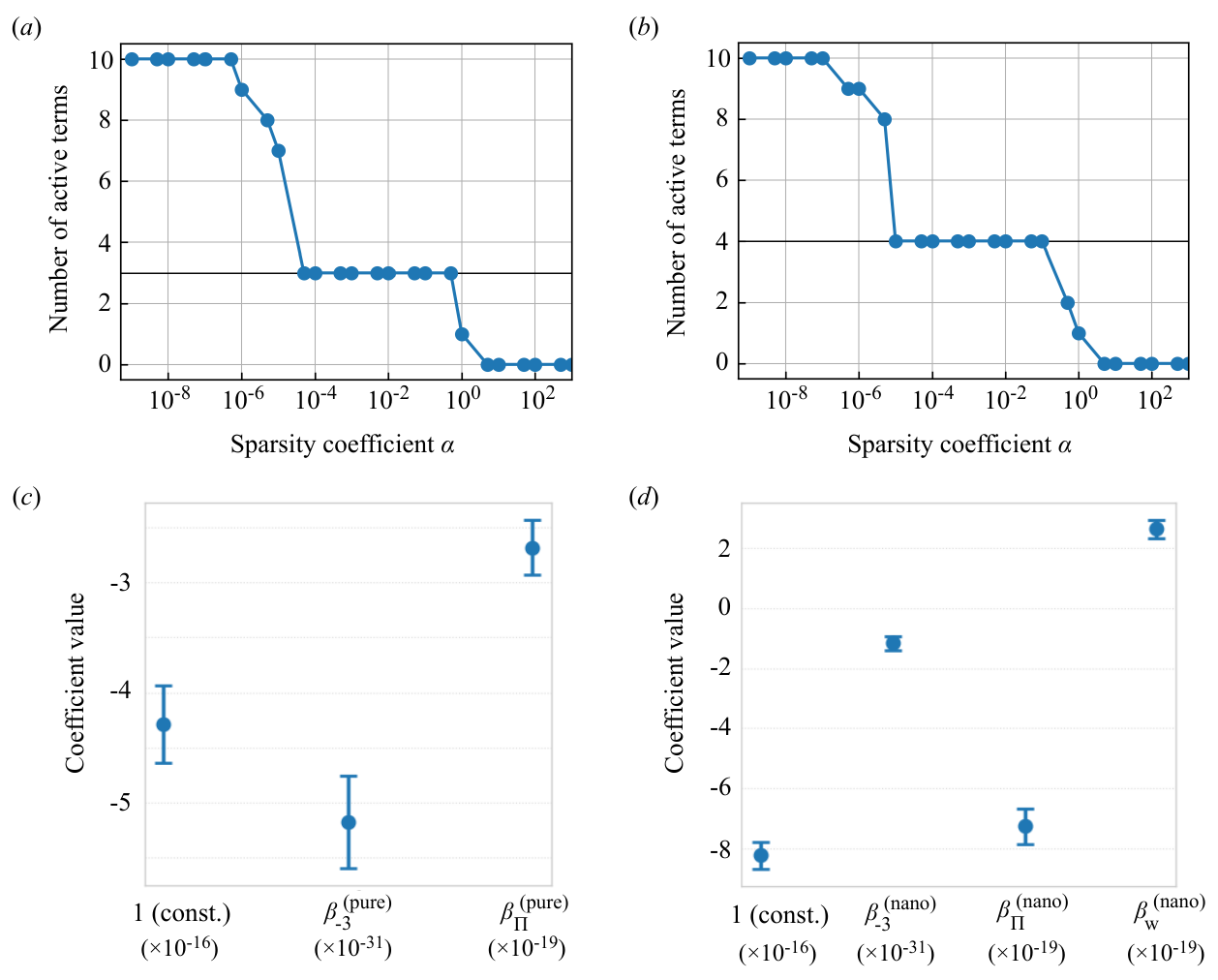}
    \caption{{
    Dependence of the number of active terms in the model identification on the sparsity coefficient $\alpha$ for $(a)$ the pure and $(b)$ nano fluids.
    The coefficients of each identified term with uncertainty for $5\times 10^{-5} \leq \alpha \leq 5 \times 10^{-1}$ in the case of $(c)$ the pure and $(d)$ nano fluids are reported.
    }}
    \label{fig:SINDyUQ}
\end{figure}

{
Let us apply the current data-driven technique to the time-resolved thickness profiles of the nanoliquid film near the contact line to infer the local dynamics of the liquid-film thickness $h(r,t)$.
The present approach finds the following sparse model forms for the pure $n$-heptane and the nanofluid:
\\
\noindent\textbf{Pure $n$-heptane:}
\begin{equation}
\label{eq:sparse-modeling-pure}
h_t
= -\,4.29\times 10^{-16}
\;-\;5.18\times 10^{-31}\,h^{-3}
\;-\;2.68\times 10^{-19}\,\big(h^{-1} h_r\big)_r
\end{equation}
\noindent\textbf{Nanofluid (surface modified CeO$_2$ in $n$-heptane):}
\begin{equation}
\label{eq:sparse-modeling-nano}
h_t
= -\,8.23\times 10^{-16}
\;-\;1.17\times 10^{-31}\,h^{-3}
\;-\;7.26\times 10^{-19}\,\big(h^{-1} h_r\big)_r
\;+\;2.63\times 10^{-19}\,\big(h^{-1}\big)_r
\end{equation}
These two models share three terms of the constant, $h^{-3}$, and $(h^{-1}h_r)_r$, while one additional term of $(h^{-1})_r$ emerges for the nanofluid.
This suggests that the $(h^{-1})_r$ term presents the effect of nanofluid injection, and we find that it is likely true from the comparison to the classical thin-film model, which is detailed later.
{The current sparsity-promoting technique almost uniquely determines these active terms, while automatically removing redundant or unnecessary terms through sparsification when the active terms are included in the library candidates.
While the current model reproduces the spatiotemporal pattern of both pure and nano fluids as depicted in figure~\ref{fig:rec}, the model cannot find the solution that provides accurate reconstruction of the time derivative, exhibiting an underfitted model, when these active terms are not included in the library.}
Furthermore, this study reveals that the order of the identified coefficients provides physical insights on the time-varying characteristics of the superspreading-wetting dynamics of a droplet.
{
The reported coefficients are averaged over a certain range of the sparsity coefficients for $5\times 10^{-5} \leq \alpha \leq 5 \times 10^{-1}$ in which the present algorithm stably provides solutions with small uncertainties in model identification, as depicted in figure~\ref{fig:SINDyUQ}.
}
}

{Note that the original measurements without pre-processing are used for the current data-driven analysis.
We have confirmed in our preliminary analysis that the identified result becomes almost equivalent even if high-frequency noise is removed from the current experimental measurements. 
This is likely because the current algorithm is a supervised learning technique that automatically extracts the physical characteristics from the data based on correlation~\cite{FFT2020,koshikawa2026convolutional}.
In turn, it can be argued that such uncorrelated experimental noise is neglected through the identification process, which is also evident from the reconstruction around $r=2~{\rm mm}$ in figure~\ref{fig:rec}$(a)$.}

{\subsection{Physical interpretation of the common operators}}

{
Let us focus on the terms that commonly appear in the identified models.
The constant term can be viewed as the effect of near-saturated evaporation and condensation, based on a well-known planar lubrication thin-film model \cite{oron1997-rmp},
\begin{equation}
\label{eq:thinfilm}
h_t
=
\;-\;\frac{1}{\rho_\ell}
\Big(
  j_0 + \Lambda p
\Big)+\partial_r\!\left(\frac{h^3}{3\mu}\,p_r\right),
\qquad
p=\Pi(h)-\gamma\,\kappa,\quad
\kappa\simeq h_{rr},
\end{equation}
where $\mu$ is the liquid viscosity, $\rho_\ell$ is the liquid density, $\gamma$ is the liquid-air surface tension, $\kappa$ is the mean curvature (approximated by $h_{rr}$ for small slopes), $j_0$ is a nearly uniform residual vapour flux set by the slight undersaturation, and $\Lambda$ is an effective pressure-evaporation coefficient. 
The disjoining pressure is expressed in the nonretarded van der Waals form $\Pi(h) = -A/(6\pi h^3)$, with Hamaker constant $A<0$ for the present Si/SiO$_2$/\(n\)-heptane/air system.
Comparison with this theoretical model reveals that the identified equations capture the phase change dynamics through two distinct source/sink terms: a constant term and a pressure-dependent term proportional to $h^{-3}$.

First, the constant term, denoted here as $\beta_{0}$, captures the spatially uniform thinning rate $-j_0/\rho_\ell$.
The identified values of $\beta_{0} \sim \mathcal{O}(10^{-16})\,\mathrm{m \cdot s^{-1}}$ represent the background evaporation rate due to slight undersaturation in the sealed cell.
Integrated over the experimental timescale ($10^4$ s), this rate corresponds to a total thickness loss of only $\sim 10^{-12}$ m, which is negligible compared to the nanometric film thickness.
This confirms the high fidelity of the environmental control, allowing us to treat the dynamics as effectively conservative with respect to the ambient atmosphere.

Second, the term proportional to $h^{-3}$ (coefficient $\beta_{-3}$) captures the local modulation of evaporation by disjoining pressure.
For the lubrication model in equation~\ref{eq:thinfilm}, this effect corresponds to the source term $-(\Lambda/\rho_\ell)\Pi(h)$.
The coefficient $\Lambda$ is physically linked to the accommodation coefficient $\alpha$ via the linearized Hertz--Knudsen relation for interface-limited kinetics, $J_{\rm evap} \approx \alpha (\rho_v/\rho_\ell) \sqrt{M/(2\pi RT)} \, \Pi(h)$.
Matching the identified coefficient $\beta_{-3}$ with this theoretical form implies an effective accommodation factor of $\alpha \sim 10^{-2}$.
Although the intrinsic accommodation coefficient for pure alkanes is typically close to unity, this lower effective value captures the combined effects of near-interface vapor diffusion resistance and the reduction of active surface area due to high nanoparticle concentration.
The clear separation of these two evaporative modes, a negligible background rate and a physically reasonable pressure-coupled response, validates the physical fidelity of the sparse modeling.
}

\begin{figure}
    \centering
    \includegraphics[width=1\textwidth]{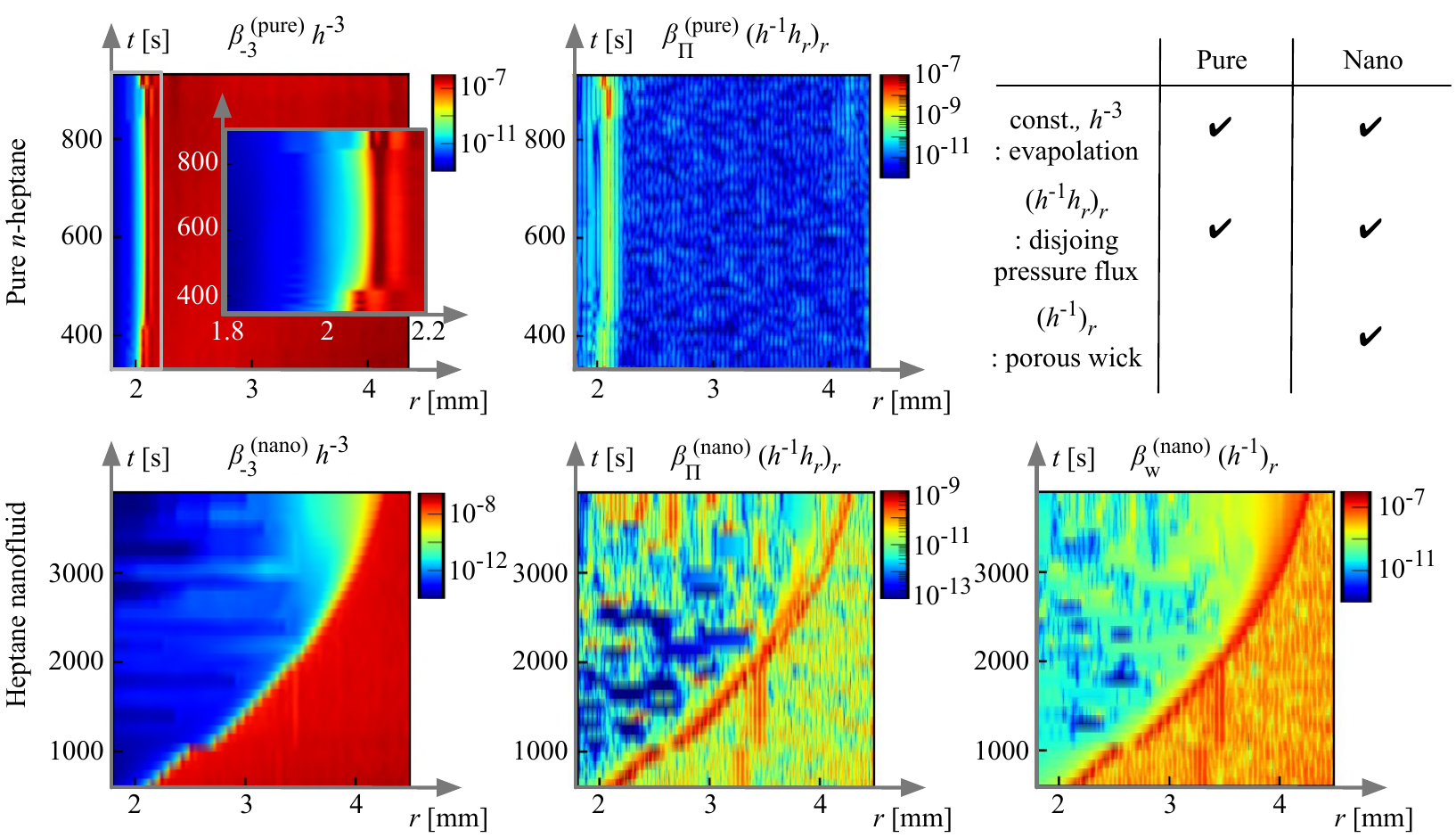}
    \vspace{-5mm}
    \caption{
    Data-driven decomposition of the droplet-height dynamics with a pure $n-$heptane and the heptane 5~wt\% nanofluid.
    The absolute logarithmic value of each term in the identified equations is presented.}
    \vspace{-3mm}
    \label{fig4}
\end{figure}

\begin{figure}
    \centering
    \includegraphics[width=1\textwidth]{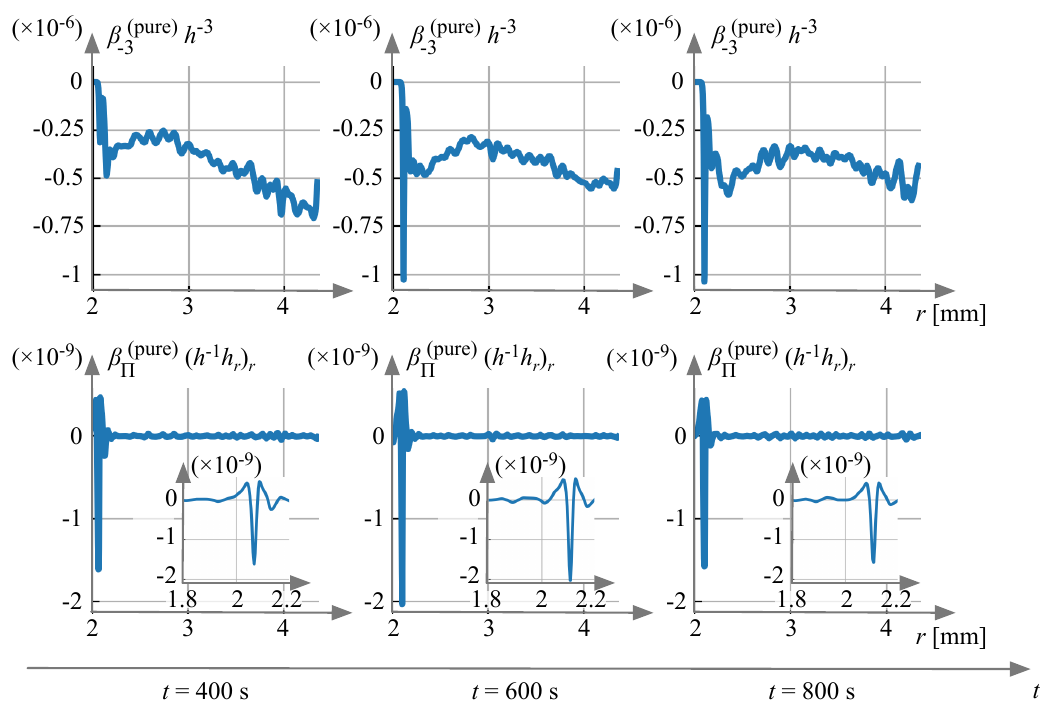}
    \vspace{-5mm}
    \caption{
    Time-varying contribution of the terms identified through the data-driven sparse modeling for the droplet height with a pure $n-$heptane.}
    \vspace{-3mm}
    \label{fig5}
\end{figure}

{
Furthermore, the term proportional to $(h^{-1}h_r)_r$, corresponding to the hydrodynamic flux driven by disjoining pressure, also appears in both of the identified models, as shown in equations~\ref{eq:sparse-modeling-pure} and \ref{eq:sparse-modeling-nano}.
We find that the order of the coefficient makes sense from the viewpoint of droplet physics, while this term nicely highlights the difference between the pure and nano fluids as that in magnitude of the coefficient.
Regarding the order of the coefficient, let us consider non-retarded van der Waals forces $\Pi(h) = -A/(6\pi h^3)$, providing the estimate of the coefficient as $\beta_\Pi = A/(6\pi\mu)$.
For the Si/SiO$_2$/\(n\)-heptane system, Lifshitz theory \cite{israelachvili2011intermolecular} predicts a Hamaker constant range of $A \simeq -(10^{-21}\text{--}10^{-20})\,\mathrm{J}$.
Given the viscosity of \(n\)-heptane, this implies a theoretical coefficient range of $\beta_\Pi^{\rm(theory)} \sim -(10^{-19}\text{--}10^{-18})~\mathrm{m^3\,s^{-1}}$.
Our sparse modeling yields $\beta_\Pi^{\rm(pure)} \approx -2.7\times10^{-19}\,\mathrm{m^3\,s^{-1}}$ and $\beta_\Pi^{\rm(nano)} \approx -7.3\times10^{-19}\,\mathrm{m^3\,s^{-1}}$, both falling within this theoretical regime.
These values correspond to effective Hamaker constants of $A_{\rm eff}^{\rm(pure)}\approx-2.1\times10^{-21}\,\mathrm{J}$ and $A_{\rm eff}^{\rm(nano)}\approx-5.6\times10^{-21}\,\mathrm{J}$.
Furthermore, the increment in magnitude, $|A_{\rm eff}^{\rm(nano)}| > |A_{\rm eff}^{\rm(pure)}|$, is physically consistent with the near-wall enrichment of CeO$_2$ nanoparticles.
Since CeO$_2$ possesses a higher refractive index and dielectric constant than both $n$-heptane and SiO$_2$, its presence enhances the effective polarizability of the film, thereby strengthening the van der Waals attraction.
}

{\subsection{Nanofluid-specific bias flux}}
{
Let us now address the unique operator $(h^{-1})_r$ identified only in the nanofluid equation.
This term represents a flux $J_{\rm bias} \propto h^{-1}$, implying a transport mechanism that becomes increasingly effective as the film thins.
Our previous measurements have found that nanoparticles persist within the nanometric precursor film \cite{shoji2024-langmuir}.
To explain the physical origin of the $(h^{-1})_r$ term, we hypothesize that these particles locally accumulate near the contact line due to evaporation, forming a porous-like structure that exerts a wicking action.
In this scenario, the rim generates a capillary suction $P_c \sim 2\gamma/r_{\rm eff}$ and possesses a permeability $k$, creating a Darcy flux that drags the adjacent liquid film.

To evaluate the plausibility of this hypothesis, we estimate the theoretical magnitude of the transport coefficient $\beta_{\rm w}$ using the scaling $\beta_{\rm w}^{\rm(theory)} \sim e (k H_{\rm p}/\mu) P_c$, where $H_{\rm p}$ is the rim thickness and $e$ is an efficiency factor.
Given the uncertainty in the microscopic structure, we evaluate the theoretical range using a rim thickness $H_{\rm p} \in [10, 50]\,\mathrm{nm}$ and porosity $\varepsilon \in [0.1, 0.5]$.
With the experimental properties ($d_{\rm p} \simeq 6\,\mathrm{nm}$, $\gamma \simeq 20.2\,\mathrm{mN\,m^{-1}}$), the ideal wicking capacity ($k H_{\rm p} P_c / \mu$) spans a window of $\mathcal{O}(10^{-19})\text{--}\mathcal{O}(10^{-16})\,\mathrm{m^3\,s^{-1}}$.
The identified coefficient $\beta_{\rm w} \approx 2.6 \times 10^{-19}\,\mathrm{m^3\,s^{-1}}$ falls near the lower bound of this physically plausible range.
This corresponds to an efficiency factor $e \sim 10^{-2}$--$10^{-1}$.
Given the complexity of the coupled transport phenomena, including evaporation, nanoparticle migration, and the resulting local variations in the rim structure, this order of magnitude renders the hypothesis physically plausible.
However, we emphasize that the capillary wicking model represents one possible physical interpretation of the identified $(h^{-1})_r$ flux.
While the quantitative agreement offered by this model is compelling, the data-driven identification of the $(h^{-1})_r$ term does not preclude alternative mechanisms that might yield similar scaling laws.
Nevertheless, the fact that the identified coefficient lies within a theoretically explainable range supports the hypothesis that nanoparticle-induced structural effects contribute significantly to superspreading, warranting further mechanistic investigation.
}

{\subsection{Data-driven decomposition of local film dynamics}}
{
To examine the dynamical balance of the identified mechanisms, let us present in figure~\ref{fig4} the magnitude of the local contribution from each operator.
Here, we consider the global spatiotemporal evolution of the absolute values of these terms (log-scale) in the $r-t$ plane.
The nanofluid significantly promotes the expansion speed and range in the $r$ direction as the contact line is clearly observed, driven by all the identified terms.
The pressure-coupled evaporation $\beta_{-3} h^{-3}$ contributes primarily in the thinnest regions, scaling naturally with film thickness.
The disjoining flux $\beta_{\Pi} (h^{-1} h_r)_r$ exhibits high intensity specifically within the precursor film region throughout the spreading process, consistent with classical precursor film theory.
Notably, the bias flux in the nanofluid, $\beta^{\rm (nano)}_{\rm w} (h^{-1})_r$, also displays elevated values in this same spatial domain.
The current observation that this term is active precisely where the precursor film exists is consistent with the hypothesis that nanoparticles within the nanoliquid film play a governing role in the superspreading dynamics.
}

{
Detailed spatial profiles for the pure $n$-heptane are shown in figure~\ref{fig5}.
Under the present saturated condition, a pre-existing adsorbed film ($\sim 10\,\mathrm{nm}$) is present ahead of the spreading front.
As the macroscopic spreading gradually arrests (at $t=600$ and $800$~s), the pressure-coupled evaporation term $\beta_{-3} h^{-3}$ exhibits a distinct, localized region of high negative intensity (large absolute value) at the transition zone ($r \approx 2.1\,\mathrm{mm}$) connecting the precursor to the adsorbed film.
This indicates significant local thinning driven by evaporation, suggesting that this localized evaporative sink effectively pins the contact line, halting the expansion.
Note that fluctuations in this term at larger radii merely reflect minor thickness variations in the adsorbed layer and lack physical significance.
Concurrently, the disjoining pressure flux $\beta_{\Pi}^{\rm (pure)} (h^{-1} h_r)_r$ displays a characteristic spatial variation within this same transition region: as $r$ increases, it exhibits a positive peak, followed by a prominent negative trough and a secondary positive peak, before asymptotically decaying to zero.
This complex oscillatory structure likely arises from the intricate mass transport required to compensate for the strong local evaporation identified above.
While the similarity of these profiles across different time steps is intriguing, a detailed stability analysis of this quasi-steady state remains a subject for future work.
}

\begin{figure}
    \centering
    \includegraphics[width=\textwidth]{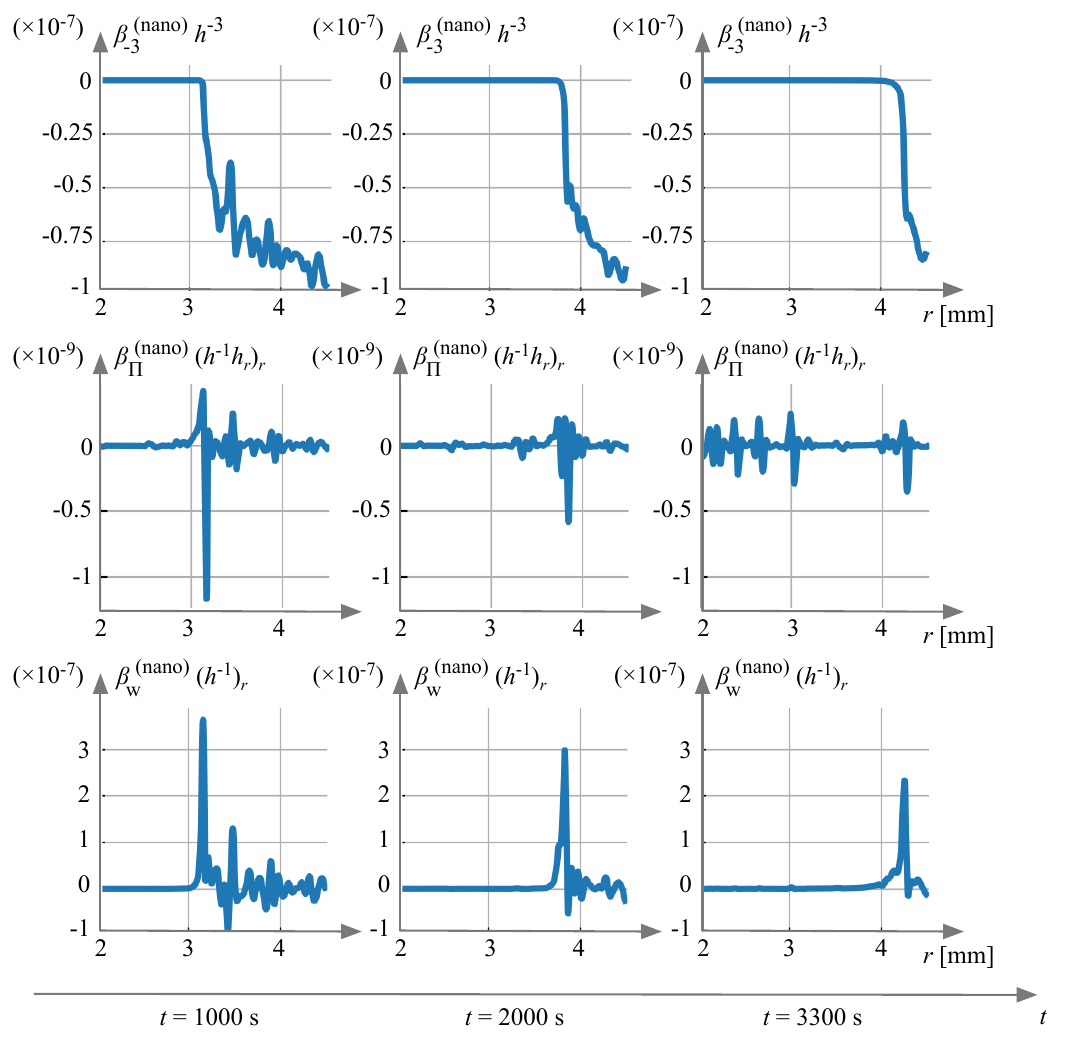}
    \vspace{-4mm}
    \caption{
    Time-varying contribution of the terms identified through the data-driven sparse modeling for the droplet height with the heptane 5~wt\% nanofluid.}
    \label{fig6}
\end{figure}

In contrast, the nanofluid dynamics, detailed in figure~\ref{fig6}, exhibit a fundamental shift in the force balance.
The pressure-coupled evaporation $\beta_{-3} h^{-3}$ follows a spatial distribution naturally scaling with film thickness, though its magnitude is approximately an order of magnitude lower than that of the pure $n$-heptane.
The disjoining pressure flux $\beta_{\Pi}^{\rm (nano)} (h^{-1} h_r)_r$ retains a spatial structure similar to the pure case, showing elevated absolute values at the advancing transition zone connecting to the adsorbed film.
However, its overall contribution is smaller than in the pure liquid.
Instead, the dynamics are overwhelmingly governed by the nanofluid-specific bias term $\beta^{\rm (nano)}_{\rm w} (h^{-1})_r$.
Unlike the complex oscillatory profile of the disjoining pressure, this term exhibits a simple, prominent positive ridge at the connection region.
Crucially, its magnitude reaches $\mathcal{O}(10^{-7})$, exceeding the disjoining pressure contribution ($\mathcal{O}(10^{-9})$) by two orders of magnitude, which identifies this term as the dominant driver of the superspreading.
Furthermore, the peak intensity of this bias flux decays over time, consistent with the experimental observation that the superspreading velocity gradually attenuates towards the later stages.
This indicates that the nanoparticle-induced transport effectively overrides the classical lubrication dynamics.

\section{Conclusions}
\label{sec:conc}

This study successfully demonstrated the capability of a data-driven sparse modeling approach, PDE-FIND, to extract the governing partial differential equations of nanoscale liquid-film dynamics directly from high-precision experimental data. By analyzing spatiotemporal film-thickness profiles acquired using phase-shifting imaging ellipsometry, we identified the distinct physical operators driving the spreading of a pure solvent and a surfactant-free nanofluid. For the pure solvent, the framework recovered classical lubrication physics driven by disjoining pressure and evaporation. In contrast, for the nanofluid, the algorithm uncovered a unique transport term scaling with the gradient of the inverse film thickness, $(h^{-1})_r$. This discovery provides a clear mathematical signature for the anomalous superspreading behavior and suggests a previously unidentified nanoparticle-induced bias flux active within the precursor film.

These findings represent a significant conceptual advance in wetting phenomena. 
Traditionally, the mechanisms of complex wetting have been investigated either by fitting macroscopic parameters to empirical scaling laws or through purely theoretical and computational deductions \cite{degennes1985,oron1997-rmp, sefiane2008-acis, bonn2009-rmp, lu2016-acis, lohse2022-arfm}. Furthermore, past discussions on superspreading origins have relied on assuming specific physicochemical mechanisms, such as Marangoni flows \cite{nikolov2002-acis}, structural disjoining pressure \cite{wasan2003-nature}, or surfactant bilayer formations \cite{karapetsas2011-jfm, theodorakis2015-langmuir}.
Our methodology fundamentally shifts this paradigm by enabling the inductive discovery of governing equations directly from nanometer-resolved local kinematics. 
Importantly, the mathematical isolation of this bias flux {supports the possible role of} nanoparticle presence within the nanometric precursor film in dictating macroscopic spreading, aligning with our recent experimental insights \cite{shoji2025-jcis,shoji2024-langmuir,shoji2026-ami}. 
The driving forces captured by our discovered PDEs are inherently rooted in smaller-scale structural formations.
To interpret the physical origins of these structures, it is crucial to connect our macroscopic findings with molecular-level structure formation and affinity evaluations using molecular dynamics simulations \cite{leroy2010-jcp, leroy2015-langmuir, surblys2018-jcp, saito2021-jcp, saito2022-aip, saito2023-jcp, sato2025-jcp}. 
Such a connection between data-driven mathematical operators and atomistic mechanisms provides a versatile framework for unraveling anomalous interfacial transport, broadly applicable to other structured precursor films of polymers \cite{villette1997-pa, xu2004-prl, glynos2011-prl, schune2019-langmuir, schune2020-acs-macro}, ionic liquids \cite{shiomoto2021-langmuir}, liquid crystals \cite{tsujita2025-langmuir}, {and other transient thin-film systems with dynamically evolving morphologies \cite{lombardi2025-jcis-1,lombardi2025-jcis-2}}.
{Several limitations should also be noted.
The present analysis was performed for a specific surfactant-free 5~wt\% CeO$_2$ nanofluid in $n$-heptane, and the structural stability of the identified nanofluid-specific term with respect to particle concentration, particle material, particle size, and surface modification remains to be examined.
Therefore, the capillary-wicking interpretation proposed here should be regarded as a physically plausible hypothesis consistent with the identified $(h^{-1})_r$ flux, rather than a definitive proof of a unique microscopic mechanism.
Future experiments using systematically varied nanofluid formulations will be necessary to determine how the coefficient and even the structure of the identified PDE depend on nanofluid properties.}

Furthermore, the extraction of governing PDEs {may provide a mechanistic basis for future continuum modeling of coating and printing processes involving nanoparticle-containing liquids}.
These advanced deposition techniques are essential for manufacturing printed electronics \cite{nayak2019-jmcc, chung2019-as}, energy devices \cite{karunakaran2019-jmca, han2020-ael}, and the fabrication of biosensors and biomaterials \cite{li2015-lc, li2020-cr, evans2021-ijp, carou2024-am}.
As these interfacial engineering technologies relying on complex wetting phenomena continue to advance across diverse industrial fields, deeply understanding the underlying physics governing droplet spreading and film evolution becomes increasingly critical.
{
Although the current model is mainly used to extract physical insights from data, the mathematical formulation discovered in this study not only facilitates direct integration into computational fluid dynamics simulations for predictive modeling but also enables the derivation of new dimensionless numbers to characterize and control complex spreading regimes.
In such a scenario, the model may need to be parameterized by including explicit variables inside the library matrix.}
The current approach could bridge the gap between fundamental nanoscale interfacial physics and scalable macroscopic applications.

\section*{Acknowledgments}

K.F. acknowledges support from the JSPS KAKENHI Grant No. JP25K23418 and No. JP26K01129, the JST PRESTO Grant No. JPMJPR25KA, and the MEXT Coordination Funds for Promoting Aerospace Utilization Grant No. JPJ000959. 
E.S. acknowledges support from the JSPS KAKENHI Grant No. JP24K01230. 
E.S. also acknowledges the Materials Processing Science Project (“Materealize”) of MEXT (Project No. JPMXP0219192801) for providing the nanoparticles used in our previous experiments.

\section*{Data availability}

The data that support the findings of this study are available from the corresponding author upon reasonable request.

\section*{Declaration of competing interest}
The authors declare that they have no known competing financial interests or personal relationships that could have appeared to influence the work reported in this paper.


\bibliographystyle{unsrt}  
\bibliography{refs}

\end{document}